# SkyWatch: A Passive Multistatic Radar Network for the Measurement of Object Position and Velocity

Mitch Randall[*,†], Alex Delacroix[†], Carson Ezell[†,‡], Ezra Kelderman[†], Sarah Little[†,§],
Abraham Loeb[†,¶], Eric Masson[†], Wesley Andrés Watters[†,‖], Richard Cloete[†,**] and Abigail White[†,‡]

[*]Ascendant AI, Boulder, CO 80304, USA

[†]Galileo Project, Harvard-Smithsonian Center for Astrophysics
60 Garden Street, Cambridge, MA 02139, USA

[‡]Harvard University, 60 Garden Street, Cambridge, MA, USA 02138

[§]Scientific Coalition for UAP Studies, Fort Myers, FL 33913, USA

[¶]Astronomy Department, Harvard University
60 Garden Street, Cambridge, MA 02138, USA

[‖]Whitin Observatory, Wellesley College
106 Central Street, Wellesley, MA 02481, USA

[**]richard.cloete@cfa.harvard.edu

Quantitative three-dimensional (3D) position and velocity estimates obtained by passive radar will assist the Galileo Project in the detection and classification of aerial objects by providing critical measurements of range, location, and kinematics. These parameters will be combined with those derived from the Project's suite of electromagnetic sensors and used to separate known aerial objects from those exhibiting anomalous kinematics. SkyWatch, a passive multistatic radar system based on commercial broadcast FM radio transmitters of opportunity, is a network of receivers spaced at geographical scales that enables estimation of the 3D position and velocity time series of objects at altitudes up to 80 km, horizontal distances up to 150 km, and at velocities to ±2 km/s (±6 Mach). The receivers are designed to collect useful data in a variety of environments varying by terrain, transmitter power, relative transmitter distance, adjacent channel strength, etc. In some cases, the direct signal from the transmitter may be large enough to be used as the reference with which the echoes are correlated. In other cases, the direct signal may be weak or absent, in which case a reference is communicated to the receiver from another network node via the internet for echo correlation. Various techniques are discussed specific to the two modes of operation and a hybrid mode. Delay and Doppler data are sent via internet to a central server where triangulation is used to deduce time series of 3D positions and velocities. A multiple receiver (multistatic) radar experiment is undergoing Phase 1 testing, with several receivers placed at various distances around the Harvard–Smithsonian Center for Astrophysics (CfA), to validate full 3D position and velocity recovery. The experimental multistatic system intermittently records raw data for later processing to aid development. The results of the multistatic experiment will inform the design of a compact, economical receiver intended for deployment in a large-scale, mass-deployed mesh network. Such a network would greatly increase the probability of detecting and recording the movements of aerial objects with anomalous kinematics suggestive of Unidentified Aerial Phenomena (UAP).

*Keywords*: Unidentified aerial phenomena; multistatic passive radar.

---

[**]Corresponding author.

## 1. Introduction

The Galileo Project (the Project) is developing and deploying a suite of field instruments in support of the scientific investigation of Unidentified Aerial Phenomena (UAP) (Watters, 2023). One of these instruments is SkyWatch, a multistatic network of passive radars based on commercial FM radio stations as transmitters of opportunity. This application of passive radar to UAP was first proposed by Davenport (1999) and later refined (Davenport, 2004). The primary measurement products produced by the instrument are bistatic range, radar cross-section (RCS), three-dimensional (3D) position, and velocity, as functions of time. These values can be derived for each one of multiple simultaneous objects within the detection volume. The SkyWatch radar detects FM-band reflective aerial objects horizon-to-horizon, at ranges up to 150 km, altitudes up to 80 km, at up to 15 times per second. The instrument is software-defined to trade off operating parameters. The SkyWatch radar complements the Galileo Project's suite of optical, infrared, acoustic, radio spectrum, magnetic field, and particle detecting instruments by providing an independent estimate of the range, location, velocity, and acceleration of multiple, simultaneous aerial objects in a comparatively large detection volume. Descriptions of the anomalous kinematics exhibited by UAP that appear to demonstrate advanced technology, including high velocities, high accelerations, abrupt maneuvers, and remaining stationary in high winds at high altitude (ODNI, 2021), make the parameters measured by SkyWatch particularly relevant to the Project.

Section 2 gives an overview of the overall network concept. In Sec. 3, we discuss the SkyWatch data products and their uses, including in conjunction with the Project's other instruments. The basic principle of operation of a single bistatic radar receiver is derived in Sec. 4. We also discuss the performance of a single radar receiver in Sec. 4. Section 5 addresses receiver performance and identifies the primary engineering challenges. The diverse geography anticipated drives the requirement for the various receiver modes discussed in Sec. 6. Section 7 describes the Phase 1 testing of the radar network being carried out at five secondary sites near the Harvard–Smithsonian Center for Astrophysics (CfA) in Cambridge, MA, USA, where the Phase 1 primary observatory-class instrument suite site is located. In Sec. 8, we discuss the next steps toward achieving the end goal of an economical, mass-deployable system with nation-scale coverage.

## 2. Overview of the SkyWatch Passive Multistatic Radar Network Concept

SkyWatch consists of a set of passive radar receivers connected and controlled by a centralized server to form a multistatic radar network, as shown in simplified form in Fig. 1. Commercial broadcast FM radio stations that transmit in the range 88–108 MHz ($\lambda = 3.4$–$2.8$ m, respectively) are used as the transmitters of opportunity. As will be detailed below, each receiver must obtain a sample of the transmit waveform (the reference signal) in order to process echo returns. This can be derived either by receiving the transmit signal directly or, if that is prevented by the geography, obtaining the reference via the internet from a receiver node closer to the transmitter.

The centralized server controls the operational parameters at each receiver node, such as their mode of operation, the reference transmitter chosen, and integration time. The server also facilitates the routing of reference signals to receivers as needed. Each receiver node computes the received power, range, and Doppler shift of the echoes it receives from objects in the airspace. That information is recorded locally and is also transmitted to the central server. The central server performs correspondence processing to relate detections in one radar receiver to detections in another. Grouped detections are then processed by an Extended Kalman Filter (EKF) to estimate 3D positions and velocities of scatterers at regular intervals in near real-time. These data are used in both real-time, for detection of objects and flagging of anomalous kinematics, and recorded for later use in combination with data from the rest of the Project's instrument suite, for more detailed object discrimination.

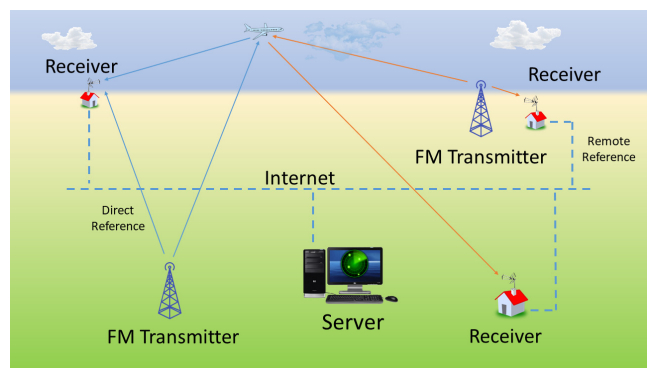

Fig. 1. (Color online) The SkyWatch passive multistatic radar network consists of two or more radar receiver nodes, one or more commercial FM broadcast transmitters of opportunity, and a centralized server. Receiver nodes are connected via the internet to a server. The server controls the operational modes of the receivers, routes reference data, and processes final triangulated results in real time.

## 3. SkyWatch Data Products and their Uses

### 3.1. *Object position*

Information about an object's 3D location in geographic space derived from SkyWatch can contribute to the Project's target acquisition system. Real-time position information can be used in conjunction with that from wide-field cameras to assist in aiming the narrow-field telescope at targets of interest. In Phase 1, a pan-tilt Beacon 8 security camera (8 MP) is used as the tracking telescope; in Phase 2, a higher resolution mirror-tracking telescope system will be added (Watters, 2023). Once an object is within the field of view of a narrow-field telescope, the camera will autonomously track the object; SkyWatch position information will then assist in maintaining tracking if the target is momentarily obscured by clouds. SkyWatch position information can also assist the wide-field camera AI algorithms to distinguish targets from noise.

Beyond assisting narrow field instruments in tracking objects, the position information provided by SkyWatch shall further assist the interpretation of any objects imaged by narrow-field instruments. The range information is crucial in allowing a single telescope's pointing angle and image data to provide primary independent measurement of physical dimensions, positions, velocity, and acceleration.

In the following, although we assume 3D positions and velocities are derived by combining outputs of multiple, non-colocated SkyWatch receivers, 3D positions and velocities can be determined in other ways. Individual receiver data can be integrated with ray-tracing from a single co-temporal, narrow-field instrument to independently deduce 3D position.

### 3.2. *Object velocity*

The SkyWatch radar 3D velocity information is unique in that it is measured directly by Doppler shift, rather than by finite difference of position. As a result, these estimates are unusually robust and accurate from low speeds up to 2 km/s (5.8 Mach), which exceeds the performance envelope of many known phenomena. Velocity estimates can serve as an input to refine the tracking of narrow-field instruments with pan-tilt servo loops. Velocity information can be used to estimate future position and thus can provide advanced alert of objects that may be on course to enter the operational range of the other instruments. A simple upper velocity threshold can also be used to indicate an object of interest whose flight characteristics mark it as anomalous, possibly representing a UAP, thus marking it out for targeting with the tracking telescope. As well, a threshold set to detect an object with zero velocity (i.e. not drifting with the wind) can provide another indication of an object of interest.

### 3.3. *Object kinematics*

Of primary utility is the instrument's ability to measure rapid time series of 3D object position and velocity (up to 15 samples per second). Such time series, referred to as tracks, are critically important for determining the kinematics that could distinguish anomalous object behavior from prosaic behavior.

The direct measurement of velocity allows the inference of acceleration by first-order finite difference, whereas second-order finite difference is required with position-only measurements resulting in much greater uncertainty. Acceleration is a key parameter that relates directly to the performance envelope of known propulsion technologies and promises to play an important role in distinguishing and quantifying an object's aerial behavior as anomalous and classifiable as a UAP. Acceleration can be estimated in real time and thresholded to serve as an alarm to other instruments of unusual activity and marking an object of interest for the tracking telescope.

### 3.4. *Object reflections*

Radar echoes in the operating frequency range of SkyWatch are to be expected from visible craft such as drones, rockets, other vehicles, but are not to be expected from visible atmospheric phenomena such as clouds and mirages (Watters, 2023). By comparing the Project's optical detections with SkyWatch's radar results in the same location, discrepancies in expected correspondence between the two can help identify anomalous conditions and objects of interest for the targeting telescope.

### 3.5. *Detection volume*

The SkyWatch object detection range of 150 km is approximately an order of magnitude greater than that of the Project's wide-field optical and IR cameras (Watters, 2023). As such, an important feature of SkyWatch is its large detection volume per receiver, even in heavy cloud cover or fog. SkyWatch-derived object positions can provide target acquisition information from beyond the operational scope of the wide-field cameras. Furthermore, a relatively sparse SkyWatch network can cover a large area, even on a national scale. The probability of detecting UAP is expected to increase roughly proportionally to detection volume (Poher & Vallee, 1975). SkyWatch's order of magnitude greater detection range over wide-field optical and infrared cameras translates to about three orders of magnitude increased probability of detecting the presumably rare UAP we seek to study.

## 4. Passive Radar Receiver Principle of Operation

Commercial FM radio stations transmit a sine wave of constant amplitude but whose frequency varies according to the content being broadcast (Federal Communications Commission, 2006). Normalized, the transmit signal $f(t)$ is represented as

$$f(t) = e^{i(\omega t + \phi(t))}, \quad (1)$$

where $\omega$ is the carrier frequency and $\phi(t)$ is the instantaneous additive phase due to the frequency modulation by the source material. For the purposes of this work we shall refer to a normalized replica of the transmit signal as the reference. The transmitted signal reaches objects in the sky and scatters in all directions. The scattered signal that is received is scaled by a constant factor $A$ in amplitude and phase, and delayed by the transit time $\delta$ from the transmitter to the object and then to the receiver. We represent this received echo $g(t)$ as

$$g(t) = Af(t - \delta) = Ae^{i((t-\delta)\omega + \phi(t-\delta))}. \quad (2)$$

We now multiply the received echo $g(t)$ by $f^*(t - \tau)$, the complex conjugate of the reference delayed by an arbitrary time $\tau$. If the reference delay $\tau$ is equal to the scattered signal delay $\delta$, the product $h(t)$ will be a coherent signal

$$h(t) = g(t)f^*(t - \tau)|_{\tau=\delta}$$
$$= Ae^{i((t-\delta)\omega + \phi(t-\delta))}e^{-i((t-\tau)\omega + \phi(t-\tau))}|_{\tau=\delta} = A. \quad (3)$$

If motion of the object gives rise to an additive received-echo path length difference $\Delta r$, and we again set the reference delay $\tau$ equal to the nominal received echo delay $\delta$, conjugate multiplication yields

$$h(t) = Ae^{i((t-\delta-\frac{1}{c}\Delta r)\omega + \phi(t-\delta-\frac{1}{c}\Delta r))}e^{-i((t-\delta)\omega + \phi(t-\delta))}$$
$$= Ae^{-i\omega\frac{1}{c}\Delta r}e^{i(\phi(t-\delta-\frac{1}{c}\Delta r) - \phi(t-\delta))}. \quad (4)$$

Provided the modulation $\phi(t)$ is relatively unchanged over a small change in delay time $\frac{1}{c}\Delta r \ll 1$ ms, the second term reduces to

$$e^{i(\phi(t-\delta-\frac{1}{c}\Delta r) - \phi(t-\delta))} \approx 1. \quad (5)$$

Thus, we see that when multiplying the conjugate of the reference delayed by $\tau$ equal to the nominal delay $\delta$ of a received echo, a small change in received echo path length results in a change in the phase of the product as

$$h(t) \approx Ae^{-i\omega\frac{1}{c}\Delta r}. \quad (6)$$

The phase of $h(t)$ advances by $2\pi$ for each change in wavelength ($\lambda = \frac{2\pi c}{\omega} \approx 3\,\text{m}$) of received echo path length difference $\Delta r$. In this way, the product formed with the conjugate of the reference delayed by the nominal delay of the received echo of a scatterer in motion produces a corresponding Doppler signature.

We now introduce the concept of bistatic range defined as the echo path length from transmitter to scatterer to receiver minus the direct path length from transmitter to receiver $R_t + R_r - R_d$. With this definition, the minimum bistatic range of any scatterer is zero, corresponding to a scatterer located somewhere along the line between the transmitter and the receiver.

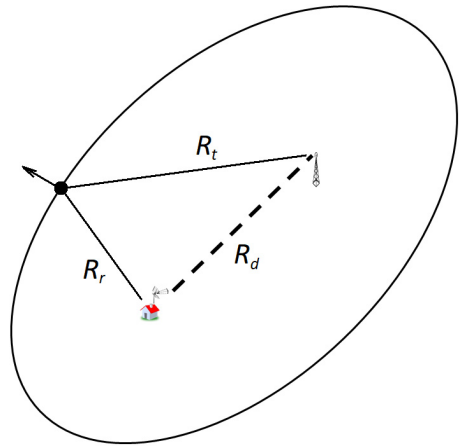

Fig. 2. (Color online) A surface of constant bistatic range forms an ellipsoid (an ellipse in cross-section). A change in path length represents a scatterer motion perpendicular to this surface.

A scatterer of a particular positive bistatic range lies on an ellipsoidal surface as shown in cross-section in Fig. 2. The Doppler frequency of received echoes relates to a constant change in the bistatic range with time; an object velocity component in the direction normal to the ellipsoidal surface of constant bistatic range.

If the reference delay does not match the transit delay ($\tau \neq \delta$), the signal is

$$h(t) = g(t)f^*(t-\tau)$$
$$= Ae^{i((t-\delta)\omega + \phi(t-\delta))}e^{-i((t-\tau)\omega + \phi(t-\tau))} \quad (7)$$
$$= Ae^{i(\tau-\delta)\omega}e^{i(\phi(t-\delta)-\phi(t-\tau))}. \quad (8)$$

We make a change of variables $\gamma = \tau - \delta$, and integrate over an interval $T$

$$\hat{h} = \frac{Ae^{i\gamma\omega}}{T}\int_a^{a+T} e^{i\phi(t)}e^{-i\phi(t-\gamma)}dt. \quad (9)$$

For typical FM radio content, $e^{i\phi(t)}$ behaves much like a unit vector with uniformly distributed random angle (Lauri *et al.*, 2007). Thus, its statistics average like a random walk (Pathria, 1995) so that

$$|\hat{h}| \approx \frac{|A|}{\sqrt{T}}. \quad (10)$$

Suppose that a linear combination of two echoes of amplitude $A_1$ and $A_2$ and of different delays is present in the received signal and we form the product with the conjugate of the reference at a delay $\tau$ that matches the nominal delay $\delta_1$ of the first echo. The resulting product $h(t)$ will contain the linear combination of a coherent signal $A_1 e^{-i\omega\frac{1}{c}\Delta r}$ resulting from the first echo, plus a noise-like signal of amplitude $A_2$ from the second echo. We can integrate $h(t)$ over an interval $T$ in order to reduce the magnitude of the noise-like signal to $\frac{|A_2|}{\sqrt{T}}$. The greater the integration interval, the greater the ability to distinguish echoes of differing delays expressed in decibels as $10\log_{10}(N)$, where $N$ is the number of independent samples integrated. Herein, we refer to the process of conjugate multiplication followed by integration as correlation at delay $\tau$.

We construct a practical realization of our correlation process by multiplying the received echo by a delayed reference and passing the result through a low-pass filter (Hall, 1995) as shown in Fig. 3. For each delay $\tau$, we can determine the received amplitude and Doppler shift corresponding to the implied bistatic range. Assuming a sampled data system, software can step through a span of delays from zero to some maximum and determine the amplitude and Doppler shifts for each of several delays. When triangulated with other receivers or located along a ray by an imager such as a camera, these delays correspond to object ranges and the Doppler frequency shifts correspond to their velocities.

## 5. Receiver Performance

We now wish to explore receiver sensitivity. To compute the received power from a scatterer in the bistatic case, we break the usual radar equation (Skolnik, 2008) into two factors as

$$P_r = \frac{P_t G_t G_r \lambda^2 \sigma}{(4\pi)^3 R_t^2 R_r^2} = \frac{P_t G_t G_r \lambda^2}{(4\pi)^2 R_t^2}\frac{\sigma}{4\pi R_r^2}, \quad (11)$$

where $R_t$ is the distance from the transmitter to the scatterer, $R_r$ is the distance from the receiver to the scatterer, $P_r$ is the received power, $P_t$ is the transmitted power, $G_t$ and $G_r$ are the transmit and receive antenna gains, respectively, $\lambda$ is the wavelength, and $\sigma$ is the RCS (Knott *et al.*, 1993) in square meters. We recognize the first factor as the Friis transmission loss over a distance $R_t$ (Stutzman & Thiele, 1998), and the second factor as the geometric loss of an ideal isotropic scatterer over a distance $R_r$. Using the value of 15 dBsm for the RCS of a typical single-engine light aircraft (Patriarche *et al.*, 1976), a transmitter Effective Isotropic Radiated Power (EIRP) = $P_t G_t$ of 100 kW, a wavelength of 3 m, a receive antenna gain of 2 dBi, and a distance $R_t = R_r = 55$ km, we compute an expected receive signal power of $-116$ dBm. Setting $R_t = R_r = 55$ km defines the Signal-to-Noise Ratio (SNR) for the general case regardless of the exact configuration of transmitter, scatterer, and receiver. Assuming a receiver noise figure of 1 dB and a noise bandwidth of 200 kHz, the thermal noise floor is about −120 dBm. Using digital averaging in the receiver to reduce the sample rate to 15 samples/second would afford about 41 dB of processing gain, so that the resulting SNR of the detected signal would be around 45 dB. As we shall see, this favorable result is available only in special cases where the direct transmit signal is blocked by terrain or structures. In practice, sample rates above 15 samples/second are generally impractical, because the associated integration time is too short to result in an acceptable output SNR.

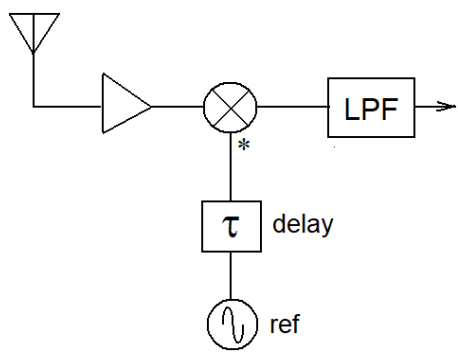

Fig. 3. (Color online) Correlative range processing where the received signal is multiplied by the complex conjugate of a replica of the transmit signal delayed by time $\tau$. The product is passed through a low-pass filter to perform the averaging needed to sufficiently resolve echoes in bistatic range.

A takeaway from the derivations above regarding correlation is that an echo corresponding to a particular delay $\delta \neq \tau$ will be randomized into a noise-like signal when correlated with the reference as delayed by $\tau$. In many cases, this noise-like signal can be many times greater than the receiver's thermal noise floor. The most troublesome example of a signal of this kind comes directly from the transmitter to the receiver. The received power of the direct signal can be estimated by the Friis transmission loss formula

$$P_r = \frac{P_t G_t G_r \lambda^2}{(4\pi)^2 R_d^2}, \qquad (12)$$

which is just the first factor of Eq. (11) where $R_d$ is the direct distance from the receiver to the transmitter. In this case, the expected receive power is about −25 dBm. When this signal is correlated against the reference at any other delay, its phase will be randomized into a noise-like signal about 95 dB greater than the thermal noise floor. To generalize this result, we see by inspection that when the three distances $R_t, R_r, R_d$, are of the same order, the ratio of the amplitude of the echo signal and the direct signal is the second factor of Eq. (11):

$$\frac{\sigma}{4\pi R_r^2} \qquad (13)$$

and is typically in the range of about −90 dB. Thus, careful attention must be paid to mitigate ground clutter returns and the direct-path signal so that the desired signals can be adequately recovered from their noise-like interference. For environments where the transmitter of opportunity is nearby, and not blocked by terrain, the mitigation of this signal is the primary engineering challenge of the radar receiver.

If we assume that a 10 dB SNR is sufficient, we can work backwards to determine the largest acceptable amount of power the radar can cope with from the direct signal. We do this for the case at hand as a point of reference. Assuming the echo signal power is −116 dBm, we see the noise floor must be no greater than −126 dBm after the processing gain (averaging). For a processing gain of 41 dB, this means the direct signal must not be greater than −85 dBm. If we were to average for 1 s, the processing gain would be 53 dB, corresponding to a tolerable direct signal power of −73 dBm. This means that the direct signal power of −25 dBm must be attenuated by 48 dB. Greater attenuation is required to achieve the same SNR with higher sampling rate.

A prototype of a single SkyWatch receiver was constructed and made to operate based on a direct reference as per the work of Howland *et al.* (2005). The prototype system uses two antennas. A first antenna is arranged to provide a null substantially directed towards an area of interest, Denver International Airport (DIA), while directing its main lobe toward the transmitter. A second antenna is arranged to direct a reception null at the transmitter while its main lobe is directed toward DIA. The orientation of the antennas for the prototype system was manipulated by mechanical adjustment. The combination of the manually adjusted antennas and the adaptive noise canceller described by Howland *et al.* (2005) sufficiently mitigated the direct signal so that the system could simultaneously receive dozens of echoes, presumably from aircraft, out to 120 km and greater.

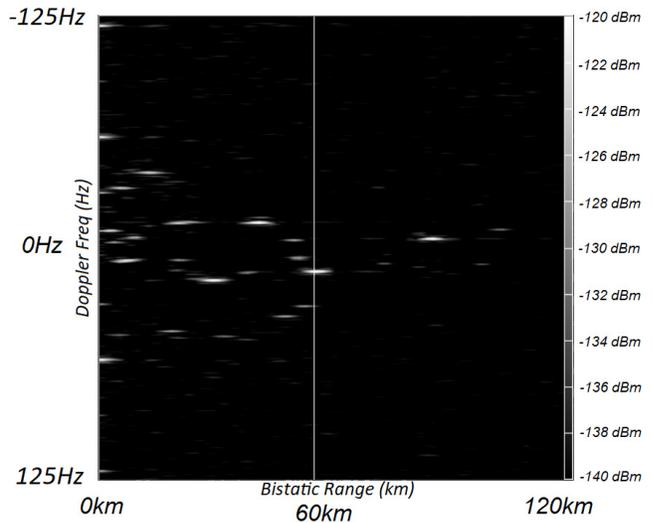

Fig. 4. ARD plot of detections from the prototype SkyWatch receiver located in Boulder Colorado. Detections at the far right indicate objects as far as 120 km in bistatic range.

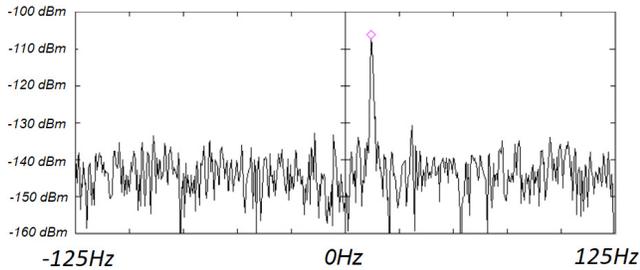

Fig. 5. (Color online) Frequency spectra of the correlation at gate 37 ($\tau = 185\,\mu s$) showing an echo amplitude of $-108\,\text{dBm}$ with a Doppler offset of 24.4 Hz corresponding to the vertical line in Fig. 4.

Each column of pixels in the Amplitude Range Doppler (ARD) plot of Fig. 4 will represent a spectrograph of the low-pass filter output of Fig. 3 computed with unique reference delay $\tau$. A slice of the ARD indicated by the vertical white line of Fig. 4 at a bistatic range of 55 km is shown as the spectrograph of Fig. 5 to demonstrate quantitatively the nature of the processed returns. As can be seen, the echo received from an airplane is about 30 dB above the noise floor with narrow spectral width.

The removal of the direct signal from the receiver has proven to be the greatest engineering challenge associated with the direct reference reception mode. The two primary methods of direct signal mitigation are digital adaptive antenna null steering (Zhenwei, 1984), and adaptive noise canceling. Both of these methods are computationally intensive, to the point of affecting system complexity and cost.

## 6. Receiver Operational Modes

There are two primary operational modes dictated by site and transmitter geography. In one mode, the reference is derived by directly receiving the transmit signal. In a second mode, the reference is provided via internet by another receiver located closer to the transmitter. These modes of operation are required to accommodate the variety of situations expected to be encountered in the mass deployment of a network of receivers.

In the context of mass deployment for mesh systems, it is expected that site selection will more often be dictated by site availability rather than by system performance or desired operation mode. For any given situation, there remain a few system parameters that can be varied in order to optimize results. A first choice is the transmitter of opportunity to be used. It is possible the site could be near an FM transmitter that allows operation in the direct reference mode. Alternatively, there may be no FM station near the site that can support the direct reference mode, in which case a remote reference can be provided from another node via the internet. A third possibility is that the site is located where the receiver could operate in either mode depending on which FM transmitter of opportunity is used. The geography that influences the modes of operation also affects data collection. When the transmitter power is attenuated by the curvature of the Earth, its ability to illuminate low altitude objects is correspondingly diminished. Deployment in a valley can be challenging, because transmitters would likely be located high on the surrounding ridge, giving rise to unfavorable ground clutter issues. In light of the number of choices at hand, the SkyWatch receivers will allow remote control of their variable operation parameters. The centralized server can then find the best configuration by simply cataloging the performance of several combinations of FM transmitters received in the various modes.

### 6.1. *Direct reference mode*

When the power from the direct signal of the transmitter is greater than about 30 dB above the thermal noise floor, the receiver may operate by receiving the direct signal and creating a reference from it as demonstrated by Howland *et al.* (2005). The method relies heavily upon a means of canceling the reference signal from the echo signal using a joint Gradient Adaptive Lattice (GAL) predictor and Normalized Least Means Square (NLMS) adaptive noise canceling filter. The method requires the use of multiple antennas in order to create statistically independent channels (Haykin, 2002), as well as to provide additional attenuation of the direct signal from the transmitter in the echo channel. Digital adaptive null steering is used to maximize the attenuation of the direct signal in the echo channel (Zhang *et al.*, 2011).

### 6.2. *Remote reference mode*

When the power from the direct signal of the transmitter is less than about 30 dB above the thermal noise floor, the receiver may operate by using a reference transmitted from a remote monitor as was demonstrated by the Manastash Ridge passive radar in Washington by Sahr & Lind (1997) and Hansen (1994). In this method, a monitor receiver is located near the transmit antenna and receives the transmit signal with good fidelity. The data is compressed and sent in time-stamped packets via internet to the echo receiver system where range gates are correlatively processed for echo detection. In this mode, there is no need for digital antenna null steering nor adaptive noise canceling. However, this mode requires precise time and frequency synchronization between nodes. The SkyWatch receivers achieve this using GPS disciplined oscillators and 1 pps GPS time stamps marking the packets.

### 6.3. *Hybrid reference mode*

When the direct signal of the transmitter is around 30 dB above the thermal noise floor, the direct signal may be too contaminated with multipath interference to be useful as a reference, and yet the direct signal may be strong enough to limit the detection of smaller objects. The two modes discussed above present very different but not mutually exclusive engineering and design challenges that allow a hybrid mode. In this case, a combination of the two modes can be used to achieve acceptable performance. In this case, a remote reference is used to perform the correlations, but digital antenna null steering and adaptive noise canceling are used to limit the strength of the direct signal to the extent possible. Without good SNR on the reference signal, a great deal of cancellation cannot be expected of adaptive noise cancellation. However, every 1 dB of direct signal cancellation ultimately translates to 1 dB of SNR improvement. So, even 20 dB of direct signal cancellation can bring the system to acceptable performance levels.

### 7. Phase 1 Multistatic Experiment

Several experimental SkyWatch receivers as shown in Fig. 6 have been installed at secondary sites around the observatory system at the CfA, where the full Galileo Project Phase 1 instrument suite is located as part of a proof-of-concept study. The SkyWatch receivers are designed to use commercial FM broadcast stations as transmitters of opportunity. They employ receivers and vertical dipole antennas capable of receiving signals in the commercial FM band from 88 MHz to 108 MHz at a sample rate of up to 1 Msps. This testing phase will continue for several months. There are many objectives for this phase, including performance validation, testing and evaluation of operation and data processing algorithms, assessment of the utility of the data product, and gaining experience that will inform the design of a future mass-deployable version. This section describes a subset of these objectives, opportunities for development, and areas for improvement and study.

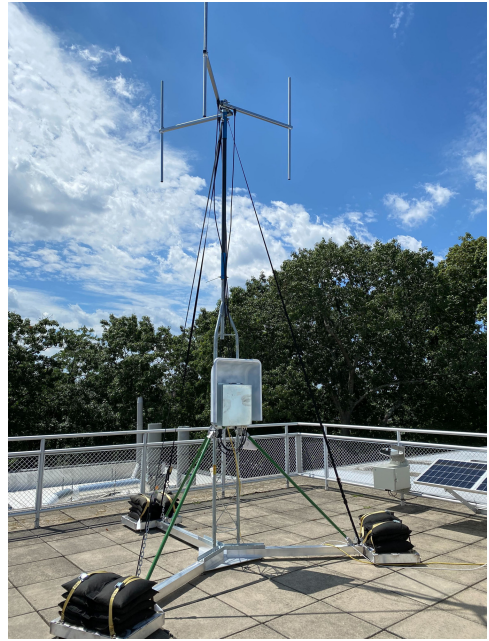

Fig. 6. (Color online) Initial prototype SkyWatch receiver deployed in the initial multistatic experiment. A three-element, vertically polarized, circular dipole array is used for adaptive null steering and direction-of-arrival interferometry.

### 7.1. *Geographic configuration*

The system as currently envisioned requires receivers spaced on a geographic scale well outside the perimeter of the primary instrument site at CfA in order to retrieve 3D positions and velocities. Figure 7 shows the placement of six receiver sites and the three chosen transmitters of opportunity. Selection of these sites was based on requirements for power, internet, security, as well as the willingness of cooperative land owners to host sites. Four of the sites are within 30 km of the primary installation while two are located much further (103 and 148 km), nearer distant transmitters.

Three transmitters of opportunity are considered for this experiment: a first transmitter A with an EIRP of 50 kW located generally to the south, a second transmitter B transmitting an EIRP of

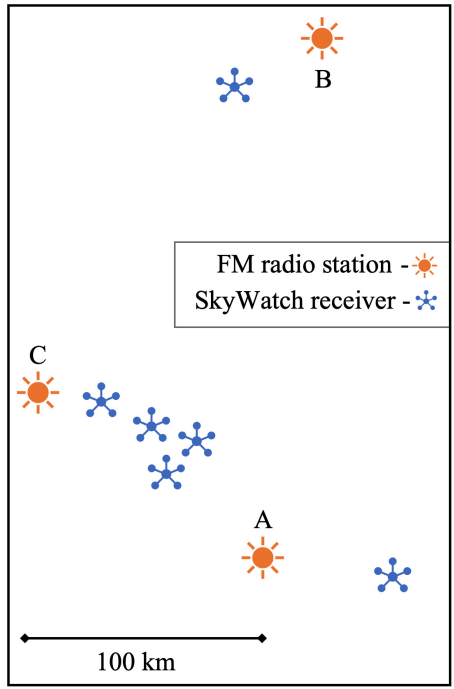

Fig. 7. (Color online) Geographic configuration of Phase 1 SkyWatch receiver deployment showing the relative placement of six radar receiver locations and three FM radio transmitter sites.

100 kW located generally towards the north, and a third transmitter C with an EIRP of 37 kW located generally towards the west. Various combinations of receivers and transmitters can be formed with this configuration. Further, receivers can act as remote reference receivers to support operation of other receivers in remote reference mode. Transmitters at a range of distances and transmit powers were chosen to allow study of pairing permutations. The latitude, longitude, and elevation of receivers is determined using GPS. The latitude, longitude, and elevation of transmitter antennas is determined using satellite imagery and Federal Communications Commission (FCC) data.

### 7.2. *Data collection*

For the Phase 1 system, all receivers intermittently record unprocessed samples to disk. All of the receivers follow the same predetermined schedule of recording for 300 s every 1024 s (about 17 min) synchronized to the same GPS time epochs. The receive frequency is stepped through a list of the three transmitter frequencies at each recording interval so that all receivers are receiving the same frequency at the same time. Each interval generates one file per receiver. With three receiver channels, at the sample rate of 1 Msps and 8-bit in-phase and quadrature (complex) samples, this amounts to a file size of about 1.8 GB per receiver every 17 min. Each receiver is equipped with a 5G telecommunications modem allowing it to establish a high-speed data connection to a server. A background task syncs the local files to the server and then removes them locally. The recording of raw data was chosen as the preferred method for the purpose of development. In this mode, post processing can be done to pair any receiver with any transmitter, and operate in any reference mode, allowing A/B comparisons of operating modes and algorithms with invaluable real-world data. Each data file contains a header indicating several operational parameters such as the Unix time epoch corresponding to the beginning of data collection, receiver latitude, longitude, and elevation as determined by GPS, receive frequency, sample rate, gain setting, and descriptive name. Data file names include the receiver latitude, longitude, frequency, date, time, and file type to assist in post-processing data management and organization.

### 7.3. *Receiver performance study*

An engineering objective of this experiment is to better understand receiver performance in real-world geography. Several parameters will be studied to understand better how they trade off with performance. One of the most consequential trade-offs relates to the reference mode, whether direct or remote. A receiver that is geographically shielded from the direct transmit signal could potentially have extraordinary sensitivity as well as design simplicity in remote reference mode. However, that comes at the expense of potentially poor illumination of low-altitude objects. On the other hand, a receiver operating in direct reference mode does not suffer from this problem, but at the cost of considerably more computational complexity as well as a much larger antenna.

### 7.4. *Adaptive beam forming*

To support economy and mass deployment, an approach using electronic beam steering rather than manual or mechanical antenna steering is employed. Each receiver is equipped with a three-element, circular, vertical dipole array. Digital beam forming is used to synthesize the two required independent apertures needed to satisfy the conditions required for adaptive noise cancellation (Haykin, 2002) as used in direct reception mode. In addition, an antenna pattern with a sharp null can be synthesized to mitigate the direct signal as much as possible (Zhenwei, 1984; Zhang *et al.*, 2011). Beam forming is achieved as a weighted sum of the signals received by each antenna.

$$X_1 = \sum_{i=1}^{3} w_i x_i, \quad (14)$$

where the $x_i$ are the three received signals, $w_i$ are three complex weights, and $X_1$ is the received signal of the synthesized beam. Figure 8 shows a simulated steerable null formed with a three element array. The recorded data allow beam forming and adaptive null steering to be performed in post processing.

### 7.5. *Tracking algorithm development*

Observations of an object by multiple receivers can be used to determine object position and velocity by triangulation. This is done most directly by use of the EKF (Sorenson, 1985). The EKF is the ideal tool in this application because it provides an optimal estimation with awareness of measurement errors, and is forgiving of intermittent or overdetermined data. It further does not require explicitly solving the geometric triangulation problem. Any method of triangulation, however, requires the association of individual detections of one receiver with the corresponding individual detections of another receiver. This is a challenging problem for SkyWatch as its detection volume is so large, and by nature it sees all detectable objects in this volume simultaneously. This is a nontrivial problem and the choice of the best approach to solve it is an area of current development. Candidates include the Kuhn–Munkres (Hungarian) algorithm (Kuhn, 1955), Multiple Hypothesis Tracking (MHT) (Blackman, 2004), Probability Hypothesis Density (PHD) tracking (Lin *et al.*, 2004), and algorithms that employ AI such as DeepSort (Wojke *et al.*, 2017). The trade-offs and applicability of these approaches can be studied using the data collected during this experiment. It is noteworthy that echo amplitude and interferometric azimuth angle may aid in resolving the echo associations (see below).

### 7.6. *Triangulation with narrow field instruments*

Full triangulation can be achieved with one pan-tilt camera and as few as one SkyWatch receiver. A narrow-field pan-tilt camera accurately estimates a ray in the direction of the object it is tracking. However, the camera alone cannot determine the range to the object. On the other hand, a single SkyWatch receiver detects the range of many objects simultaneously, but cannot alone determine their direction. A finite list of range hypotheses can be produced by finding the intersections between the object direction ray and the set of ellipsoidal bistatic range isosurfaces associated with the objects detected by a single SkyWatch receiver. Further, the Doppler shift associated with each detected object implies a velocity for each range hypothesis. An area of study is to develop an algorithm to determine the corresponding echo and thus locate the object in three-space. This is a 1D analog to the 3D association algorithms required for SkyWatch-only triangulation, and is expected to be commensurately less complex.

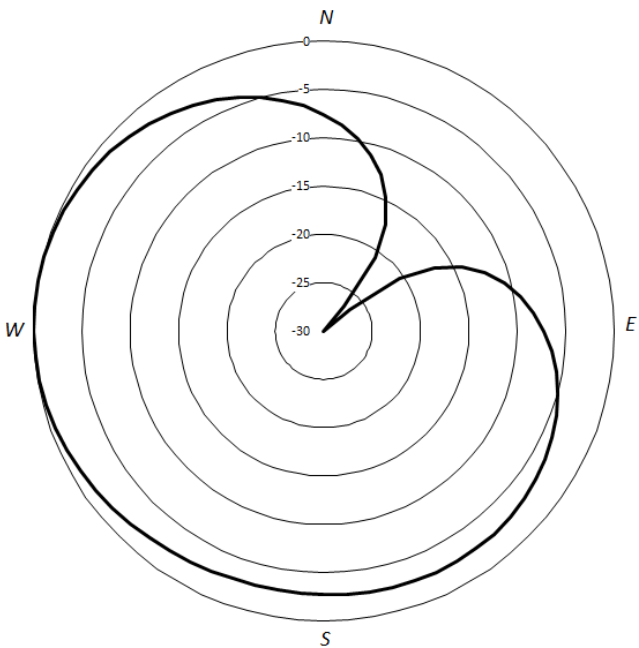

Fig. 8. Plan-view polar plot of simulated null steering with a three element array. The rings mark amplitude in 5 dB steps. The polar angle represent azimuth as viewed from above. A null can be synthesized to point in any azimuthal direction by adjustment of three complex weight factors.

### 7.7. *Radar cross-section estimation*

A further product of the radar is RCS estimation. We can use the measured receiver power to estimate

the RCS as

$$\sigma = \frac{P_r(4\pi)^3 R_t^2 R_r^2}{P_t G_t G_r \lambda^2}. \quad (15)$$

The transmitter antenna gain can depend on object position and altitude, adding uncertainty to the RCS estimate.

Typical FM transmitters employ linear antenna arrays that shape the beam toward the horizon (Shively Labs, 2006; Federal Communications Commission, 2010) as shown in the elevation pattern of Fig. 9. The RCS estimate can be improved by employing gain maps in both azimuth and elevation from the transmitter. Scattering angle can also play a role (Kenyon & Dogaru, 2017). Despite these challenges, if the RCS can be estimated to reasonable accuracy (i.e. ±5 dB), this could be used for direct comparison between receivers for the purposes of associating echoes as needed for the tracking algorithms described above. Note that radar minimum detectable RCS depends not only on distance, but also on elevation angle to the transmitter antenna as described by a curve of Fig. 9 with its peaks and nulls. The receiver dipoles have a cosine elevation response that also affects minimum detectable RCS. This is not an ideal situation and introduces RCS estimation uncertainties as well as adverse position-dependent effects. For example, an object traveling at constant altitude in a trajectory over the transmit antenna from horizon to horizon will pass through detection voids, causing detection of the object to be discontinuous over time. This also prevents a simple statement as to the sensitivity of the radar, because it depends on object elevation angle, not just range. In terms of RCS, this means that a final estimate cannot be made until object position is obtained. Notwithstanding, knowledge of the transmitter and receiver antenna patterns can be used to form an expression or numerical lookup table that can be used to help MHT and PHD association algorithms. In these algorithms, a position is assumed, and used as a hypothesis in a forward manner. Only good hypotheses should result in consistent RCS estimates from receiver to receiver.

### 7.8. *Interferometry for angle of arrival estimation*

Determination of angle of arrival (AOA) of echoes would prove helpful in associating echoes of multiple receivers with one-another in MHT or PHD tracking algorithms. Such measurement could further aid in associating radar detections with objects viewed by narrow-field instruments. In cases where the network is sparsely populated so that 3D solutions are underdetermined, AOA information could potentially be used to put upper and/or lower bounds on an object's position and/or velocity. In remote reference mode, the three-element, vertically-polarized, circular, dipole array allows for azimuthal AOA estimation by interferometry (Pan *et al.*, 2015). However, in direct-reference mode, reception of weak echoes is not possible through just one dipole, because sufficient direct signal mitigation requires a null formed by the circular dipole array. Thus, AOA interferometry in direct-reference mode is a fruitful area of future research to be explored. The SkyWatch platform with its raw recorded data provides an ideal test bed upon which various techniques for AOA determination in direct-reference mode can be attempted. Results of the various attempts can be evaluated against ground truth aircraft position and velocity data provided by Automatic Dependent Surveillance–Broadcast (ADS-B) transponder (Federal Aviation Administration, 2002) data either recorded locally, or captured historically by any number of available services.

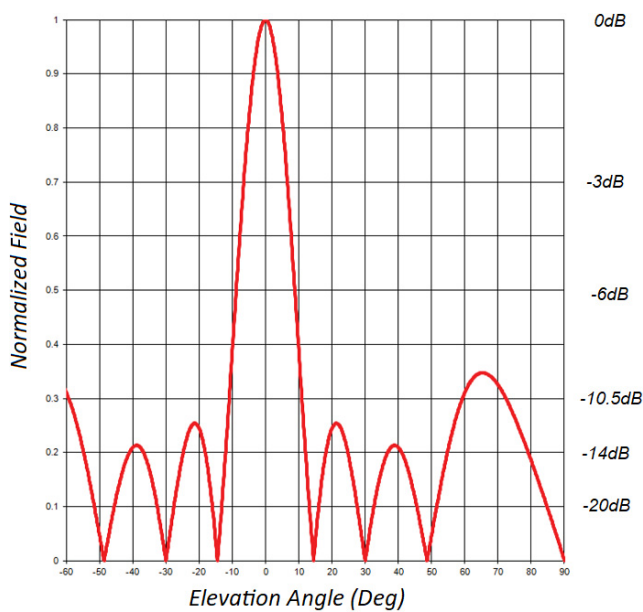

Fig. 9. (Color online) Elevation pattern of typical FM transmission antenna arrays showing main lobe directed towards the horizon. Sidelobes and nulls extend above the horizon affecting RCS estimation.

### 7.9. *Validation*

SkyWatch receivers have ample sensitivity to detect single-engine and commercial airliners, of which there can be dozens aloft within its detection volume at any one time. We exploit these detections to aid in validating position and velocity estimates obtained by the network by comparing them to known aircraft positions and velocities as provided by their ADS-B transponders. Known aircraft position can also be used to directly validate measurements from single SkyWatch receivers. These aircraft position and velocity data are readily available on the internet (FlightAware, 2017), and can also be received locally in real time (Laufer, 2018). We will also use the ADS-B aircraft type information to validate the estimated RCS. In this case, we will create a scatter chart of aircraft type versus estimated RCS to catalog the accuracy and reliability of the estimate between multiple receivers.

### 8. Next Steps

Data collected in the experiment described here will provide ample opportunities for algorithm development and system engineering. The Phase 1 SkyWatch system is designed for observatory-class systems, but the Galileo Project intends to use what it learns to develop more cost-effective passive radar systems for mass deployment in a mesh network (Watters, 2023). An important next step towards these ends is the design of a real-time version of the receiver with the goals of being smaller, less expensive, and easier to deploy. Much development is also required on the server in order to monitor and control the receiver network, as well as to interpret the data to deliver triangulated results in real-time. All of this is in service of developing an economical and truly mass-deployable receiver suitable to provide nationwide coverage (Davenport, 1999, 2004) using the many available commercial FM broadcast transmitters (the $60\,\text{dB}\mu$ service area of which is shown in Fig. 10 (Davis, 2017)) to monitor aerial objects and detect and record anomalous kinematics indicative of purported UAP. Such a system would integrate well into the small "mesh" optical, IR, and acoustic monitoring systems described in Watters (2023).

Interestingly, economy plays a nontrivial role in the quality of the science return. This is because a simple receiver solution could enable a more densely-populated network and hence better results. On the other hand, the more complex design may appear superior in its individual performance, but ultimately result in a sparsely populated network with less accurate estimates and a lower probability of detecting radar-reflective objects exhibiting anomalous kinematics that might be classified as UAP. At the nation-wide scale, pairing observatory-class SkyWatch receivers with mass-deployable versions of Galileo Project's wide-field and narrow-field optical instruments as mesh nodes has great advantages. The higher degree of corroboration would increase the certainty of measurements acquired during any captured UAP activity. Further, the greatly expanded detection volume would raise significantly the probability of detecting UAP.

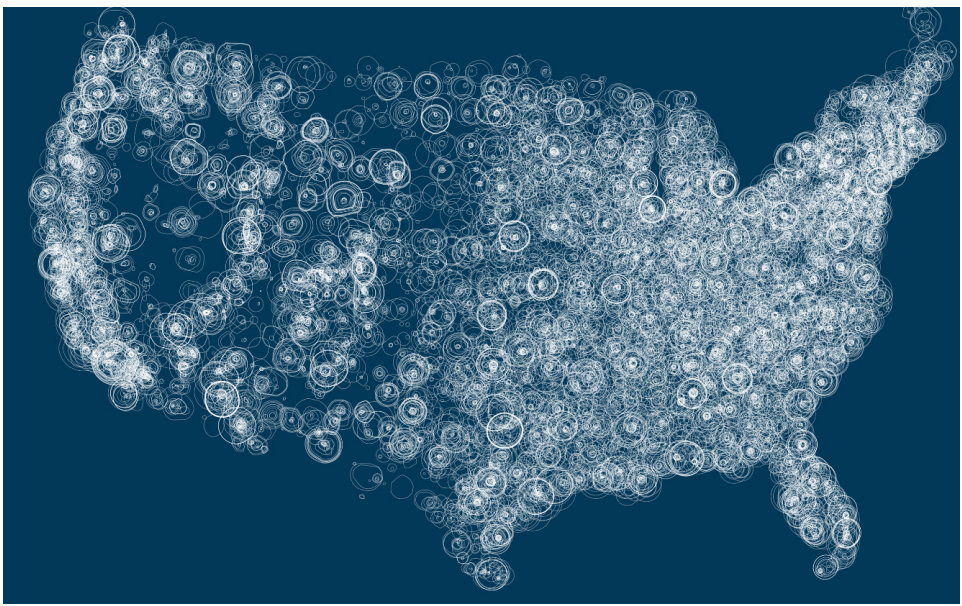

Fig. 10. (Color online) Visualization of FM broadcast radio coverage in the US. Note that useful radar operation extends significantly beyond the power contours plotted (Visualization courtesy E. Davis https://erdavis.com).

## 9. Conclusion

A network of multistatic passive radar receivers based on commercial FM transmitters of opportunity shows great promise in providing key quantitative real-time data to complement the suite of instruments being developed by the Galileo Project. Real-time position and velocity data can provide advance notice of aerial craft with anomalous kinematics that might be indicative of UAP activity and help steer narrow-field instruments to them. Rapid sampling of the hemisphere in a 150 km radius will allow unprecedented estimation of the high aerial accelerations and abruptly changing kinematics that are characteristics of particular note in UAP reports. The basic theory of operation is presented, and key engineering challenges identified. Basic passive radar receiver operation has been confirmed and reproduces the results of others. A Phase 1 design has been implemented in order to test SkyWatch receivers as a network. The experiment calls for the receivers to intermittently record their raw samples and transmit these data to a server. These data are used in post processing to test several geographical and operational variations. The resulting study will inform the engineering of a system that is hoped to be smaller, more economical, and more easily deployable. It will further aid the development of the necessary server-side software and algorithms needed to support the network and to generate object positions, velocities, and radar signatures in real time. The experiment calls for validation of derived object 3D position, velocity, and radar signature against known aircraft data as recorded from ADS-B transponder messages. The vision for SkyWatch is an economical, mass-deployable instrument fielded on a nation-wide scale that maximizes the probability of detection and characterization of UAP activity.

## Acknowledgments

Special thanks to Peter Davenport of the National UFO Reporting Center for the vision and inspiration for the SkyWatch passive radar network concept. The author also wishes to thank Jeff Keeler, Paul Howland, and Darek Maksimiuk for their help and support in understanding the issues and achieving first results. Many thanks go to the entire Galileo Project team for their support.